\documentclass[twocolumn,pre,aps,showpacs,amsmath,amssymb,superscriptaddress]{revtex4}

\usepackage{hyperref}
\usepackage{amsmath,amssymb}
\usepackage{amsfonts,amsthm}
\usepackage{graphics}
\usepackage{graphicx}
\usepackage{dcolumn}
\usepackage{color}
\usepackage{bm}

\usepackage[normalem]{ulem}

\begin{document}
\title{Anticipated synchronization in neuronal circuits unveiled by a phase-resetting curve analysis}

\author{Fernanda S. Matias}
\thanks{fernanda@fis.ufal.br}
\affiliation{Instituto de F\'{\i}sica, Universidade Federal de Alagoas, Macei\'{o}, Alagoas 57072-970, Brazil}
\author{Pedro V. Carelli}
\affiliation{Departamento de F\'{\i}sica, Universidade Federal de Pernambuco, Recife, Pernambuco 50670-901, Brazil}
\author{Claudio R. Mirasso}
\affiliation{Instituto de Fisica Interdisciplinar y Sistemas Complejos, IFISC (CSIC-UIB), Campus Universitat de les Illes Balears, E-07122 Palma de Mallorca, Spain}
\author{Mauro Copelli}
\affiliation{Departamento de F\'{\i}sica, Universidade Federal de Pernambuco, Recife, Pernambuco 50670-901, Brazil}

\begin{abstract}

Anticipated synchronization (AS) is a counter intuitive behavior that has been observed in several systems. 
When AS establishes in a sender-receiver configuration, the latter can predict the future dynamics of the former for certain parameter values.
In particular, in neuroscience AS was proposed to explain the apparent discrepancy between information flow and time lag in the cortical activity recorded in monkeys. 
Despite its success, a clear understanding on the mechanisms yielding AS in neuronal circuits is still missing.
Here we use the well-known phase-resetting-curve (PRC) approach to study the prototypical sender-receiver-interneuron neuronal motif.
Our aim is to better understand how the transitions between delayed to anticipated synchronization and anticipated synchronization  to phase-drift regimes occur.
We construct a map based on the PRC method to predict the phase-locking regimes and their stability.
We find that a PRC function of two variables, accounting simultaneously for the inputs from sender and interneuron into the receiver, is essential to reproduce the numerical results obtained using a Hodgkin-Huxley model for the neurons.
On the contrary, the typical approximation that considers a sum of two independent single-variable PRCs fails for intermediate to high values of the inhibitory connectivity between interneuron.
In particular, it looses the delayed-synchronization to anticipated-synchronization transition.
\end{abstract}
\maketitle

\section{Introduction}

Anticipated synchronization (AS), as proposed originally by H. Voss~\cite{Voss00,Voss01a}, is a particular kind of lag synchronization that can occur in two unidirectionally-coupled dynamical systems (Sender-Receiver) when the receiver is subject to a self-inhibitory feedback loop.
In the counterintuitive AS regime, the receiver system can predict the future dynamics of the sender for certain parameter values.
Anticipated synchronization has been found both experimentally and numerically in different fields including optics~\cite{Masoller01b,Liu02,Tang03}, electronic circuits~\cite{Voss02}, neuronal systems~\cite{Ciszak03,Toral03b,Matias11,Pyragiene13,Matias14,Matias15,Matias16} and more~\cite{Kostur05,Pyragas08,Ciszak15,Voss16}.
In neuronal systems, AS was originally studied numerically by Ciszak and coworkers~\cite{Ciszak03,Toral03b} using a Fitzhugh-Nagumo model with diffusive coupling.
Chemical synapses in a three-neuron sender-receiver-interneuron (SRI) motif were included by Matias et al.~\cite{Matias11}. 
Using the Hodgkin-Huxley model, a transition from the more intuitive delayed-synchronization (DS) regime to the AS regime was found when changing the inhibitory conductance impinging on the receiver neuron.  
Recently, the ideas introduced  in~\cite{Matias11} were extended to neuronal populations ~\cite{Matias14} to explain the observations of a positive Granger Causality, with well-defined directional influence, accompanied by either a positive or negative phase lag in the recordings of the motor cortex activity of monkeys while doing a visual task ~\cite{Brovelli04,Salazar12}.
Despite the interest attracted by AS in neuronal circuits, a thorough understanding of how this particular state is stablished is missing.

In this paper we use the well-known phase resetting curve (PRC) approach to gain insight into the AS regime and in particular into the DS/AS transition that occurs in the SRI motif of model neurons shown in Fig.~\ref{fig:intervals}a.
In its pulsatile version, PRCs describe how the spiking time of an oscillating neuron is altered by synaptic inputs. 
PRCs have a long history in the analysis of coupled oscillators~\cite{PRCbook}. 
More than fifty years ago, for instance, this technique was used to understand how excitatory and inhibitory pulses could decrease or increase firing rates of pacemaker neurons~\cite{Perkel64}.
Despite its generality, the technique is particularly suitable when a neuron receives one input per cycle. 
As reviewed by Goel and Ermentrout~\cite{Goel02}, as well as by Canavier and Achuthan~\cite{Canavier10pulse,CanavierBookChapter}, pulsatile PRC was applied in some particular cases, namely, models of two uni- and bi-directionally coupled neurons, neurons arranged in a ring configuration, two-dimensional and all-to-all networks.
PRCs were also measured experimentally in different biological systems, from neurons~\cite{Galan05} to circadian rhythms~\cite{Czeisler86, Strogatz90light}.

We aim at comparing the predictions from the PRC technique with numerical simulations of the SRI motif whose nodes are described by Hodgkin-Huxley neuronal models with chemical synapses. 
The manuscript is organized as follows. 
The SRI motif and neuronal models are described in section~\ref{HH}. 
In section~\ref{results} we develop the PRC map for this system and compare the results with the numerical integration of the full model.
Finally, we summarize our results in section~\ref{conclusions}.

\section{The Sender-Receiver-Interneuron motif\label{HH}} 

In the SRI motif, the sender node S projects an excitatory synapse onto the receiver node R, which also receives an inhibitory projection from the node I. 
Moreover, the node R projects an excitatory synapse onto node I, closing an excitatory-inhibitory loop  (see Fig.~\ref{fig:intervals}a). 

Each node of the motif is described by the Hodgkin-Huxley model~\cite{HH52}, which consists of four differential equations describing the evolution of the membrane potential and the currents flowing across a patch of an axonal membrane~\cite{Koch}:
\begin{eqnarray}
C_m \frac{dV}{dt} &=& \overline{G}_{Na} m^3 h (E_{Na}-V) + \overline{G}_{K} n^4 (E_{K}-V) \nonumber \\
&& + G_m (V_{rest}-V) + I_c + \sum I_{syn} \label{eq:dvdt}\\ 
\frac{dx}{dt} &=& \alpha_x(V)(1-x) -\beta_x(V) x \label{eq:alphaHH} \;.  
\end{eqnarray}
$V$ is the membrane potential, $x\in\{h,m,n\}$ are the gating variables for sodium ($h$ and $m$) and
potassium ($n$). The capacitance of a $30\times 30\times \pi$~$\mu$m$^2$ equipotential patch of
membrane is $C_m = 9\pi$~pF~\cite{Koch}.
$E_{Na}=115$~mV, $E_{K}=-12$~mV and $V_{rest}=10.6$~mV are the reversal potentials of the 
Na$^+$, K$^+$ and leakage currents, respectively. 
The maximal conductances are $\overline{G}_{Na}=
1080\pi$~nS, $\overline{G}_{K} =324\pi$~nS and $G_m=2.7\pi$~nS,
respectively.
$I_{syn}$ accounts for the chemical synapses arriving from other neurons and $I_c$ accounts for an external constant current. 
In the absence of synapses $I_{syn}=0$ and for $I_c=280$~pA the neuron spikes with a period equals to $T=14.68$~ms. 
The voltage-dependent rate variables in the Hodgkin-Huxley model have the form:
\begin{eqnarray}%
\alpha_n(V) & = & \frac{10-V}{100(e^{(10-V)/10}-1)},  \\
\beta_n(V) & = & 0.125e^{-V/80}, \\
\alpha_m(V) & = & \frac{25-V}{10(e^{(25-V)/10}-1)},  \\
\beta_m(V) & = & 4e^{-V/18},\\
\alpha_h(V) & = & 0.07e^{-V/20}, \\
\beta_h(V) & = & \frac{1}{(e^{(30-V)/10}+1)}, \label{eq:betah}
\end{eqnarray}%
where all voltages are measured in mV and the resting potential is shifted to zero mV.

For the synapses we assumed a current-based model given by:
\begin{equation}
I_{syn}(t) = g_{syn} V_{syn} \sum_{spikes} \alpha(t-t_{spike}) \;.
\label{eq:HHcurrent}
\end{equation}
 $V_{syn}$ is taken, without loss of generality, equal to $1$~mV. $g_{syn}$ represents the
maximal conductances which are different for AMPA ($g_{exc}$) and GABA$_A$ ($g_{inh}$) mediated synapses.
The internal sum is extended over all the presynaptic spikes occurring at $t_{spike}$.

The $\alpha(t)$ function, that models the postsynaptic conductance, is described by the following equation:
\begin{equation}
\label{eq:alphafunction}
\alpha(t)= \pm \frac{1}{\tau_{-}-\tau_{+}} ( \exp{(-t/\tau_{-})} - \exp{(-t/\tau_{+})} ).
\end{equation}
The positive signal accounts for excitatory synpases whereas the negative for inhibitory ones. The parameters $\tau_{-}$ and $\tau_{+}$ stand respectively for the decay and rise time of the function and determine the duration of the synaptic response.
In the simulations we fix the maximum excitatory conductance $g_{exc}=1000$~nS, $\tau_{-}=6.0$~ms and $\tau_{+}=0.1$~ms.

\section{Results \label{results}}
\subsection{Phase map}

\begin{figure}
\begin{center}
\includegraphics[width=1.1\columnwidth,clip]{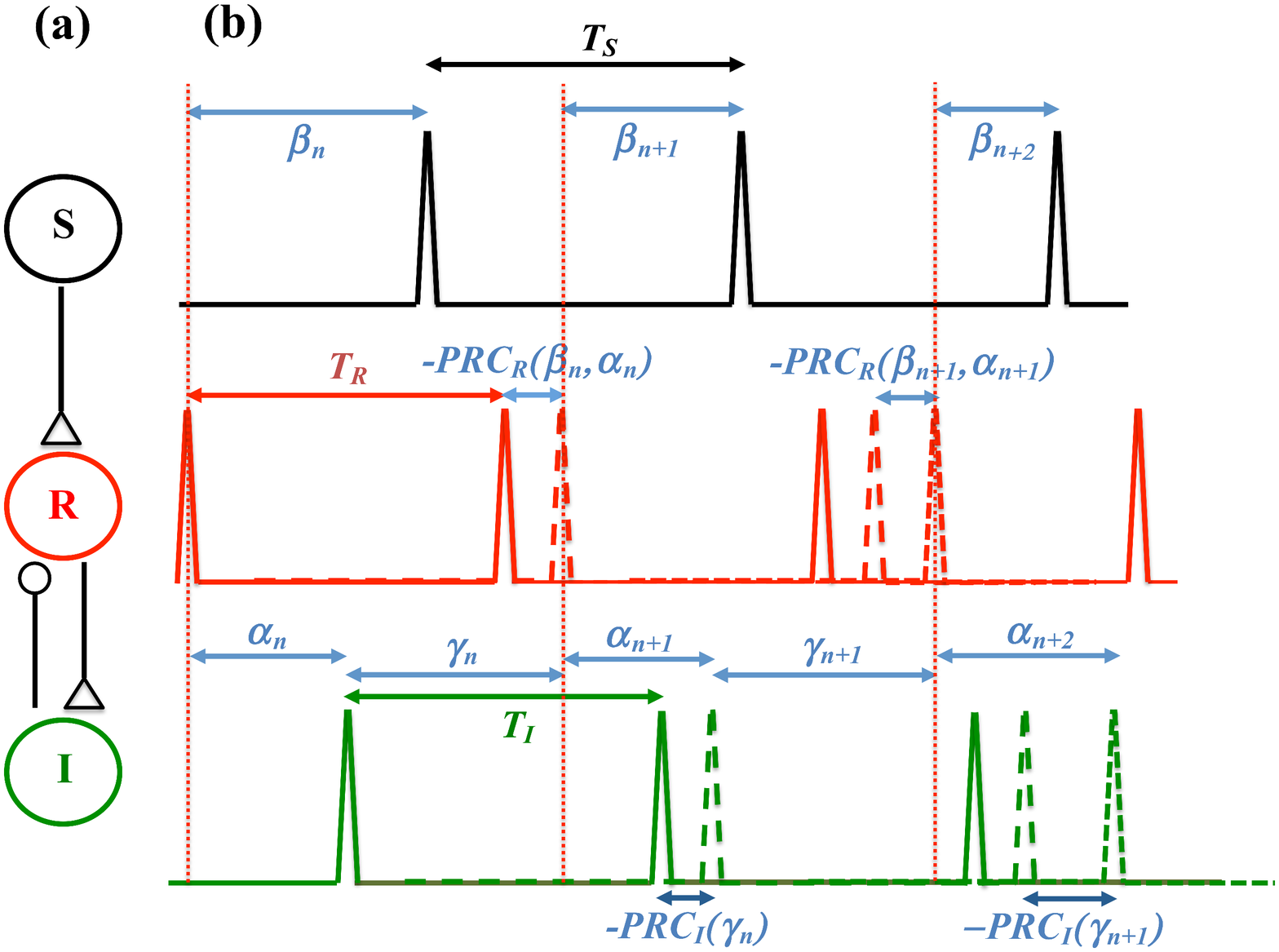}
\end{center}
\caption{
\label{fig:intervals}
a) Three coupled neurons in a SRI configuration. 
Each spike of the receiver (R) is perturbed by the synaptic 
current from the sender (S) and the interneuron (I), 
whereas each spike of the interneuron is only perturbed
by the synaptic current from the receiver. 
b) The Poincar\'e map of this configuration provides 
the time differences between the three neurons
in the phase-locking regime.
}
\end{figure}

In order to apply the PRC approach to the SRI configuration shown in Fig.~\ref{fig:intervals}, one has to consider that the central neuron receives two inputs per cycle when locked in the 1:1 solution: one excitatory (from the sender) and another inhibitory (from the interneuron). 
Following the approach initially developed in Ref.~\cite{MatiasTese}, we define $t_{R}[n]$ as the spiking time at the $n$-th cycle of the receiver, which we take as the reference to measure time differences. 
Let $t_{S}[n]$ and $t_{I}[n]$ be, respectively, the spiking times of the sender and interneuron immediately after $t_{R}[n]$ and $T_S$, $T_R$ and $T_I$ the free-running periods of the neurons, as shown in Fig.~\ref{fig:intervals}b. 
To construct the return map, we introduce the variables (depicted in Fig.~\ref{fig:intervals}b)
\begin{eqnarray}
\beta_n & \equiv &t_{S}[n] - t_{R}[n] \; ,  \nonumber \\
\gamma_n & \equiv &t_{R}[n+1] - t_{I}[n] \; ,  \label{eq:variables}
\\ 
\alpha_n & \equiv &t_{I}[n] - t_{R}[n] \; . \nonumber
\end{eqnarray}
From the above definitions, $\beta_n$ and $\alpha_n$ measure, respectively, the timing of the excitatory and inhibitory inputs relative to the receiver cycle.
$\gamma_n$ measures the timing of the excitatory input relative to the interneuron cycle. 

\begin{figure}
\begin{center}
\includegraphics[width=0.9\columnwidth,clip]{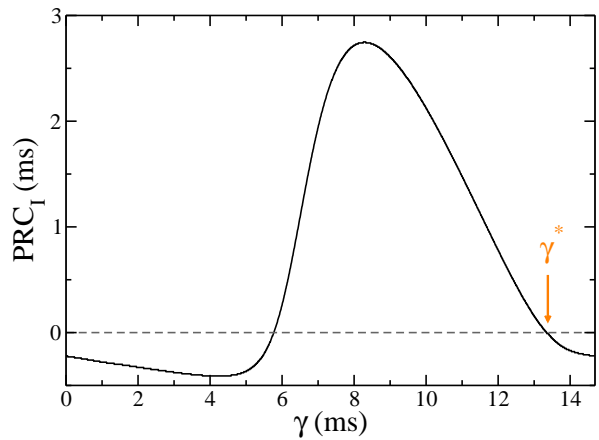}
\end{center}
\caption{Phase Resseting Curve for the interneuron as a function of $\gamma$ (see Eq.~\ref{eq:variables}).
$\gamma^*$ is the stable fixed point solution obtained with the condition given by Eqs.~\ref{eq:fixedpoints}.
}
\label{fig:singlePRC}
\end{figure}

The $PRC_x$ of a given neuron $x$ is defined as the difference between its free-running period and the period after a perturbation is applied (so that positive PRCs imply period shortening). 
We use the synaptic funtion in Eq.~\ref{eq:alphafunction} as the appiled perturbation in such a way that $PRC_x(\delta)$ is the response due to an input $\alpha[(t-\delta)$mod$(T_x)$]. 
We start by analyzing the simplest case of the interneuron, whose $PRC_I$, shown in Fig.~\ref{fig:singlePRC},  depends only on the excitatory input from the receiver.
From Fig.~\ref{fig:intervals}b we can start building the return map. 
The interval between two consecutive spikes of the I neuron satisfies
\begin{equation}
\label{eq:alpha}
T_{I} + (-PRC_I(\gamma_n)) = \gamma_n + \alpha_{n+1}\; .
\end{equation}

The analysis of the R neuron is more complicated, because it receives two inputs from different neurons at different times within one period.  
In the most general form, therefore, $PRC_R$ depends on the two variables $\alpha$ and $\beta$ that, as can be seen in Fig.~\ref{fig:intervals}b, satisfy the condition:
\begin{equation}
T_{R} + (-PRC_R(\beta_{n},\alpha_{n})) = \alpha_{n} + \gamma_{n} \; .
\end{equation}
Isolating $\gamma_n$ we get
\begin{equation}
\label{eq:gamma}
\gamma_{n} = T_{R}  - PRC_R(\beta_{n},\alpha_{n})) - \alpha_{n} \equiv \gamma_n(\beta_{n},\alpha_{n})  \; .
\end{equation}
This indicates that Eq.~\ref{eq:alpha}  can be written in terms of $\alpha_n$, $\alpha_{n+1}$ and $\beta_n$.
It is usually assumed that $PRC_R(\beta_{n},\alpha_{n})$ can be decomposed as the sum of two single-variable PRCs~\cite{Canavier10pulse}. 
We will show later that this approximation fails precisely in the region of parameter space where AS occurs. 

The interval between the $n$-th spike of the receiver and the $(n+1)$-th spike of the sender satisfies, as shown in Fig.~\ref{fig:intervals}b,
\begin{equation}
\label{eq:beta}
\beta_n + T_S = T_R + (-PRC_R(\beta_n,\alpha_n)) + \beta_{n+1} \; . 
\end{equation}

Finally, we obtain, using Eq.~\ref{eq:beta} and combining Eqs.~\ref{eq:alpha} and \ref{eq:gamma} the following two-dimensional map:
\begin{eqnarray}
\beta_{n+1} &=&  \beta_{n} + PRC_R(\beta_{n},\alpha_{n}) + T_{S} - T_{R}  
\; ,\label{eq:prcmapbeta} \\
\alpha_{n+1}&=&  \alpha_{n} +PRC_R(\beta_{n},\alpha_{n}) - PRC_I(\gamma_{n}(\beta_n,\alpha_n))  \nonumber \\
 && + T_{I} - T_{R} \; \label{eq:prcmapalpha}.
\end{eqnarray}
Two important assumptions were made here~\cite{Netoff05}. 
First, we assumed that the inputs affect only the following spike of each neuron, meaning that second order effects of the PRC are neglected~\cite{Canavier10pulse,CanavierBookChapter}. 
Second, we considered that the three neurons fire once in each cycle (which we know to be true from numerical integration of the equations~\cite{Matias11}). 

\subsection{Phase-locked solutions and stability}

To gain insight into the transition from anticipated to delayed synchronization (AS/DS), we  look for the fixed point solutions of Eqs.~\ref{eq:prcmapbeta} and  \ref{eq:prcmapalpha}. 
We start with the case where the three neurons have the same periods. 

\subsubsection{Identical free-running periods}

Assuming that the free-running periods of all three neurons are identical, $T_S = T_R = T_I = T$, the fixed point solutions $(\alpha^*,\beta^*)$ are given by
\begin{eqnarray}
 PRC_R(\beta^{*},\alpha^{*}) &=& 0  \nonumber \\
 PRC_R(\beta^{*},\alpha^{*}) - PRC_I(\gamma^{*}) &=& 0 \; , \label{eq:fixedpoints}
\end{eqnarray}
where $\gamma^* = \gamma_n(\alpha^*,\beta^*)$ as defined in Eq.~\ref{eq:gamma}.
In the phase-locking regime one therefore has $PRC_I(\gamma^{*})=0$.

The analysis of the system of equations \ref{eq:fixedpoints} can be done in two steps. 
First we find the stable fixed point solution for the one dimensional $PRC_I(\gamma^*)$ (note in Fig.~\ref{fig:singlePRC} that the curve has two fixed points, the one with negative slope being the stable one~\cite{Canavier10pulse,CanavierBookChapter}%
). 
Since Eq.~\ref{eq:alpha} implies $\alpha^* = T - \gamma^*$, the search of the zero of $PRC_R(\alpha^*,\beta^*)$ only requires the line with constant $\alpha^*$ to be scanned. 

\begin{widetext}

\begin{figure}[!th]%
\includegraphics[width=\linewidth]{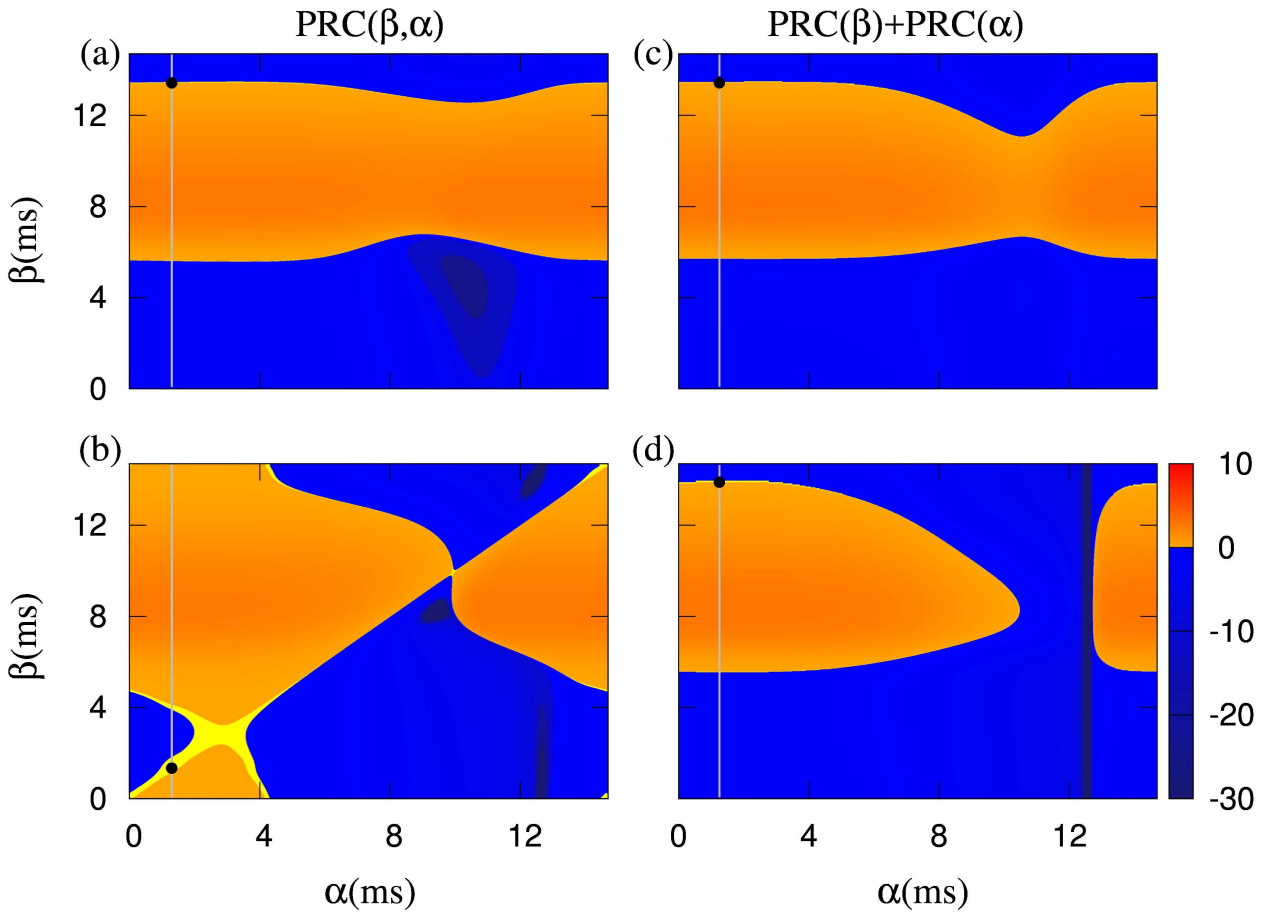}
\caption{
{\bf (Color online) Phase Response Curve of the receiver neuron due to two inputs per cycle}. 
$PRC$ is color-coded (in ms) as a function of its two variables $\alpha$ and $\beta$.  
In (a) and (b) we plot the full function $PRC_R(\beta,\alpha)$, whereas in panels (c) and (d) we plot the approximation 
$PRC_R(\alpha)+PRC_R(\beta)$. 
In the upper (lower) panels, $g_{inh}=200$ nS ($g_{inh}=1000$ nS).
The filled circles correspond to the stable fixed point, predicting delayed synchronization in panels a, c and d, and anticipated synchronization in panel b.
The prediction of panel d is incorrect (see text for details).
}
\label{fig:PRC2D} 
\end{figure}%
\end{widetext}

In Fig.~\ref{fig:PRC2D} we show $PRC_R(\beta,\alpha)$ as a function of its two arguments.
This function is obtained by numerically integrating the HH equations for the R neuron subject to one excitatory and one inhibitory inputs at different times of the R neuron period.
These two inputs are given by Eq.~\ref{eq:alphafunction}, with their appropriate parameters.
For simplicity, and without loss of generality, we keep the excitatory conductance $g_{exc}$ fixed, while we change the values of $g_{inh}$.
The points of interest in the figure are those that satisfy $PRC_R=0$.
In order to find the stable fixed point solutions of Eq.~\ref{eq:fixedpoints} we should scan the values of the PRC in a vertical line  in Figs.~\ref{fig:PRC2D} corresponding to the stable $\alpha^*=T-\gamma^*$    value shown in Fig.~\ref{fig:singlePRC}. 
The stable solution is the one that crosses zero with negative slope when increasing $\beta$ (filled circles in Figs.~\ref{fig:PRC2D}a,b). 
The line $\beta=\alpha$ is of particular importance, corresponding to the excitatory and inhibitory inputs arriving simultaneously at the receiver.
It can be clearly seen, when comparing panels b and d in Fig.~\ref{fig:PRC2D},  that the combined effect of the two pulses is very different in the full PRC function than when we just add the effect of them independently.

It is worth noting that different values of the inhibitory conductance lead to severe changes in the  $PRC_R(\beta,\alpha)$ landscape, particularly impacting the position of the fixed points relative to the pulsating period (compare Figs.~\ref{fig:PRC2D}a and b). 
This change corresponds to a transition in the synchronization regime (from DS to AS). 
Let us define the spike timing difference $\tau_{SR}$ between sender and receiver (in the phase-locking regime) as the difference between their closest spikes, i.e. $\tau_{SR}=t_{R}-t_{S}$.
Consequently, if $\beta^* < T/2$ (see Eqs.~\ref{eq:variables} and Fig.~\ref{fig:intervals}b), then the system is in a AS regime characterized by $\tau=-\beta^*$. 
On the contrary, if $\beta^* > T/2$, the system operates in the DS regime which is characterized by $\tau_{SR}=T-\beta^*$.

We now check the accuracy of the PRC prediction and compare it with the numerical simulations of the full HH model.
We show in Figs.~\ref{fig:PRC2D}a and b two examples of $PRC_R{(\beta,\alpha)}$ for two different values of the inhibitory conductance $g_{inh}$. 
For the parameters of Fig.~\ref{fig:PRC2D}a (small inhibitory conductance), the stable fixed point $\beta^*$ is clearly $>T/2$ for any value of $\alpha^*$. 
The system therefore operates in the DS regime, as long as a stable $\gamma^*$ (and consequently $\alpha^*$) exists. 

In addition, in Fig.~\ref{fig:PRC2D}c  we show the results when one employs the decomposition $PRC_R{(\beta,\alpha)} \approx  PRC_R(\alpha)+PRC_R(\beta)$.
This is the usual and simplest approximation when an oscillator receives two inputs per cycle~\cite{Canavier10pulse,CanavierBookChapter}. 
In this case, $PRC_R(\beta)$ and $PRC_R(\alpha)$ represent the phase-resetting curves of the receiver when it is subject to \textit{either} an excitatory \textit{or} an inhibitory input, respectively. 
Note that the general qualitative results of Fig.~\ref{fig:PRC2D}c are remarkably similar to those of Fig.~\ref{fig:PRC2D}a. 
Moreover, the fixed points in both figures are almost identical. 
Indeed, these results predict well the stationary spiking time difference $\tau_{SR}$ directly measured in the simulations of the full Hodgkin-Huxley motif. 

\begin{figure}
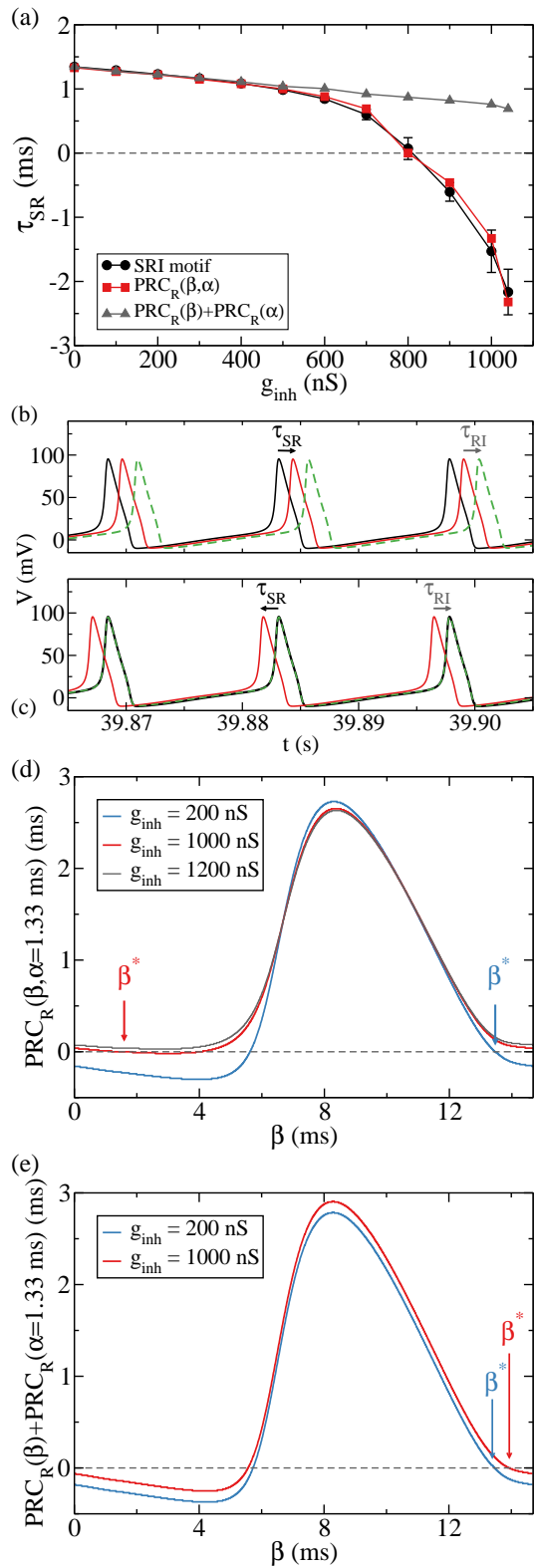

\centering
\begin{minipage}{0.9\linewidth}
\includegraphics[width=0.9\linewidth,clip]{Fig4a}
\end{minipage}
\begin{minipage}{0.9\linewidth}
\includegraphics[width=0.9\linewidth,clip]{Fig4b}
\end{minipage}
\begin{minipage}{0.9\linewidth}
\centerline{\includegraphics[width=0.9\linewidth,clip]{Fig4c}}
\end{minipage}
\begin{minipage}{0.9\linewidth}
\centerline{\includegraphics[width=0.9\linewidth,clip]{Fig4d}}
\end{minipage}
\caption{
{\bf (Color online) Comparing numerical simulation with PRC prediction. }
a) Time delay between sender and receiver as a function of the inhibitory conductance $g_{inh}$.
b) and c) time traces of the membrane potential given by Eq.~\ref{eq:dvdt} for two different $g_{inh}$ values: in b) $g_{inh}=200$ nS and in c)  $g_{inh}=1000$ nS. 
In panel b the system is locked in a delayed-synchronization regime while in panel c it is locked in the anticipated-synchronization regime.
d) The one dimensional $PRC_R(\beta,\alpha^*)$ is plotted for a fixed value of $\alpha^*=T-\gamma^*$, where $\gamma^*$ is obtained from Fig.~\ref{fig:singlePRC}, for different values of $g_{inh}$.
e) The one dimensional $PRC_R(\beta)+ PRC_R(\alpha^*)$ is plotted for the same value of $\alpha^*$ as in d. 
The difference in panels d and e reflect the discrepancies in the calculations of the fixed-point solutions for $\beta$. 
}
\label{fig:tau_gi}
\end{figure}

In Fig.~\ref{fig:tau_gi}a we plot the time difference $\tau_{SR}$ versus the inhibitory conductance $g_{inh}$. 
In the numerical simulations of the full SRI motif (full circles), delayed synchronization ($\tau_{SR} > 0$, see time traces in Fig.~\ref{fig:tau_gi}b) is obtained for $g_{inh} \lesssim 800$~nS, whereas beyond this value an anticipated synchronization regime  takes over ($\tau_{SR} < 0$, see time traces in Fig.~\ref{fig:tau_gi}c).
For $g_{inh} \gtrsim 1020$~nS, a phase drift regime is reached.
When compared with those of the PRC approach (filled squares in Fig.~\ref{fig:tau_gi}a), namely, the fixed point solutions of Eqs.~\ref{eq:fixedpoints}, results agree very well. 
The agreement extends for the whole $g_{inh}$ range, including the second transition from AS to the phase drift regime. 
Interestingly, when we approximate $PRC_R(\beta,\alpha)$ by $PRC_R(\beta)+PRC_R(\alpha)$, a good agreement is obtained only for relatively small $g_{inh}$ (filled triangles in Fig.~\ref{fig:tau_gi}a).

Why does the approximation break down? 
By examining Fig.~\ref{fig:PRC2D} it can be seen from panels b and d that the PRC landscapes are drastically different. 
In Fig.~\ref{fig:PRC2D}d, the approximation keeps the fixed point solution in the high-$\beta^*$ range, therefore predicting delayed synchronization.
This can also be seen in the projection along a constant $\alpha^*$ value shown in Fig.~\ref{fig:tau_gi}e. 
With the full $PRC_R(\beta,\alpha)$  (Fig.~\ref{fig:PRC2D}b), an increasing inhibitory conductance causes the fixed point to cross the zero-lag solution and a small value of $\beta^*$ is obtained, predicting an anticipated synchronization regime.
This transition is also illustrated in Fig.~\ref{fig:tau_gi}d. 
For even larger values of $g_{inh}$, a transition to the phase drift regime is obtained. 
Interestingly, the shape of $PRC_R(\beta,\alpha^*)$ shown in Fig.~\ref{fig:tau_gi}d for $g_{inh}=1200$~nS is reminiscent of that of a type-I excitable neuron, which is consistent with the absence of a locked regime.


\subsubsection{Different free-running periods}

Up to now, we focussed on the DS/AS transition assuming identical free-running periods for all the neurons~\cite{Matias11,Matias14,Matias16}. 
The results presented in the previous section also assumed all periods to be identical. 
However, the PRC approach, as presented in Eqs.~\ref{eq:prcmapbeta}-\ref{eq:prcmapalpha}, allows an extension to the case of different free-running periods.
Moreover, if $T_R$ does not change, the same $PRC_R(\beta,\alpha)$ shown in Fig.~\ref{fig:PRC2D}a and b is, in principle, still valid.
This, therefore, strengthens the predictive power of the PRC approach. 
To probe it, we analyze a particularly relevant scenario where the interneuron has a different period than the others. 
In neuroscience, it is often the case that inhibitory neurons spike faster than excitatory ones~\cite{Izhikevich07}. 
We therefore focus on examining the dependence of the synchronization regimes on the free-running period $T_I$ of the interneuron. 

From Eqs.~\ref{eq:prcmapbeta} and~\ref{eq:prcmapalpha} the fixed point solutions for $T_S \neq T_R \neq T_I$ become:

\begin{eqnarray}
PRC_R(\beta,\alpha) & = & \Delta T_{RS} \label{eq:PRCRDT} \\
PRC_I(\gamma; T_I) & = & \Delta T_{IS} \label{eq:PRCIDT}\; ,
\end{eqnarray}
where $\Delta T_{RS}=T_R-T_S$ and $\Delta T_{IS}=T_I-T_S$. 

Note that we have now included an explicit dependence of $PRC_I$ on $T_I$. 
To avoid recalculating $PRC_I(\gamma; T_I)$ for every $T_I$, we use instead an approximation that assumes that changes in the period amounts to a simple rescaling of the corresponding phase-resetting curve as: 

\begin{equation}
\label{eq:gammanew}
PRC_I(\gamma; T_I) = \frac{T_I}{T} PRC_I\left(\gamma \frac{T_I}{T}; T\right)\; ,
\end{equation}
where $PRC_I\left(\gamma ; T\right)$ is the function shown in Fig.~\ref{fig:singlePRC} for the case of the three neurons having identical period $T$.

The analysis of Eq.~\ref{eq:PRCIDT} with Eq.~\ref{eq:gammanew} allows discriminating two possibilities in terms of the $PRC_I$:

$\it i)$ if $T_S>T_I$ (or equivalently the sender frequency is smaller than the interneuron frequency) the fixed point solutions exist until $\Delta T_{IS}$ reaches the minimum value of $PRC_I(\gamma; T_I)$. At this value the two fixed points collide and disappear and the system enters into a phase-drift regime. 

$\it ii)$ if $T_I>T_S$ (or equivalently the sender frequency is higher than the interneuron frequency) the fixed point solutions exist until $\Delta T_{IS}$ reaches the maximum of $PRC_I(\gamma; T_I)$. At this value the two fixed points collide and disappear and the system enters into a second phase-drift regime. 

\begin{figure}
\begin{center}
\includegraphics[width=0.9\columnwidth,clip]{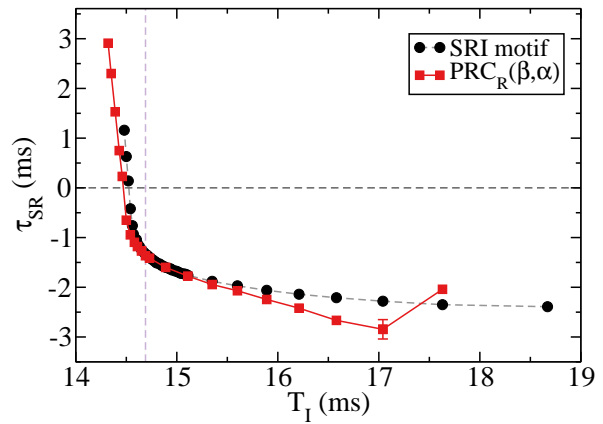}
\end{center}
\caption{
\label{fig:differentT}
{\bf (Color online) Comparison between the PRC prediction and SRI simulation for different free-running periods of the Interneuron.}
Time delay between Sender and Receiver as a function of the Interneuron period for the full simulation of the SRI motif (circles) and the PRC prediction (squares). 
The vertical dashed line corresponds to the free-running period of both the Sender and the Receiver.
The inhibitory conductance is $g_{inh} = 1000$~nS, so that the function $PRC(\alpha,\beta)$ corresponds to that of Fig.~\ref{fig:PRC2D}b. }
\end{figure}

In Fig.~\ref{fig:differentT} we plot the time difference between sender and receiver $\tau_{SR}$ as a function of the free-running period of the interneuron $T_I$. 
$\tau_{SR}$ is calculated with the fixed point solutions $\gamma^*$ obtained from Eq.~\ref{eq:gammanew} in combination with Fig.~\ref{fig:PRC2D}b.
A very good agreement can be seen when comparing the PRC's prediction with the numerical simulation of the full HH model. 
For values of  $T_I \lesssim 14.5$~ms and $T_I \gtrsim 18.7$~ms the phase-locked solution is lost and the  system enters into a phase-drift regime.

\section{Concluding remarks \label{conclusions}}

In this paper we have used a phase-reseting curve (PRC) approach to gain insight into the transition from delayed to anticipated synchronization and anticipated synchronization to phase drift regime in a sender-receiver-interneuron motif.
Initially we assumed identical parameters and operating conditions for the three neurons. 
The PRC of the receiver neuron was computed as a function of two inputs per cycle: one arriving from the sender and another from the interneuron.
We found that the description of the PRC in terms of two variables is essential to correctly match numerical results obtained from the full neuronal and synaptic model.
In particular our PRC approach correctly predicts the transition from the anticipated-synchronization to the phase-drift regime.
On the contrary, if the typical approximation is used, considering the sum of two PRCs from independent stimulus, the results significantly depart from the numerical solutions, with the largest discrepancies at intermediate to large values of the inhibitory conductance.
Moreover, this approximation does not account either for the AS/DS transition nor the AS/phase-drift regime transition observed both numerically in the full neuronal model and with the two-variable PRC. 

We have also explored the PRC prediction when the neurons had different free-running periods.
Under this condition, the PRC calculation is easily extended, in particular when only the period of the interneuron element is varied.
Assuming that the PRC of the interneuron modifies according to a simple rescaling factor when its period changes, 
we also obtain a very good agreement with the numerical simulations, highlighting the strength of the method. 
Further investigations including different types of synapses and neuronal models (type-I vs. type-II excitability) as well as different pulsating regimes will be reported in a forthcoming publication. 

\begin{acknowledgments}
We thank CNPq grant 310712/2014-9, FACEPE grant APQ-0826-1.05/15, CAPES grant PVE 88881.068077/2014-01 for financial support. 
C.R.M. acknowledges support from the Spanish Ministerio de Econom\'{\i}a y Competitividad (MINECO) and Fondo Europeo de Desarrollo Regional (FEDER) through project TEC2016-80063-C3-3-R (AEI/FEDER, UE).
This article was produced as part of the activities of FAPESP Research, Innovation and Dissemination Center for Neuromathematics (grant 2013/07699-0, S. Paulo Research Foundation).

\end{acknowledgments}
%

\bibliography{matias}

\end{document}